\documentclass[]{aa}
\usepackage{graphicx,natbib,amsmath}
\bibpunct{(}{)}{;}{a}{}{,}

\usepackage{txfonts}
\def\igr{IGR~J00291$+$5934}
\def\sax{SAX~J1808.4$-$3658}

\def\int{{\it INTEGRAL}}
\def\cha{{\textsl Chandra}}
\def\rxt{{\it RXTE}}

\def\nh{$N_{\mathrm{H}}$}

\def\ergcms{\,\text{erg}\,\text{cm}$^{-2}$\,\text{s}$^{-1}$}
\def\chis{{$\chi^2_{\nu}$}}
\def\ergs{\,\text{erg}\,\text{s}$^{-1}$}

\begin{document}
   \title{ \cha~and \rxt~spectroscopy of the accreting msec pulsar  \igr}

%   \subtitle{A study with \rxt~and \cha~}

   \author{A. Paizis\inst{1,2}, M. A. Nowak\inst{3}, J. Wilms\inst{4}, T.J-L. Courvoisier\inst{1,5}, K. Ebisawa\inst{6}, J.
   Rodriguez\inst{7,1} and P. Ubertini\inst{8}
         }

   \offprints{A. Paizis: Ada.Paizis@obs.unige.ch}

\institute{\textit{INTEGRAL} Science Data Centre, Chemin d'Ecogia 16, 1290 Versoix, Switzerland
\and
INAF-IASF, Sezione di Milano, Via Bassini 15, 20133 Milano, Italy
\and
Center for Space Research, MIT, Cambridge, MA, USA
\and
Department of Physics, University of Warwick, Coventry, CV4 7AL, UK
\and
Observatoire de Gen\`eve, 51 chemin des Mailletes, 1290 Sauverny, Switzerland
\and
NASA Goddard Space Flight Center, Code 662, Greenbelt, MD 20771, USA
\and
CEA Saclay, DSM/DAPNIA/SAp (CNRS UMR 7158 AIM) , 91191 Gif Sur Yvette, France
\and
CNR-IASF, Sezione di Roma, via del Fosso del Cavaliere 100, 00133 Roma, Italy
}

   \date{Received 13 May 2005 / Accepted 14 July 2005}

   \abstract{We report on an observation of the recently discovered
accreting millisecond X-ray pulsar \igr~performed 
with the \rxt-Proportional Counter Array (PCA) and \cha-High Energy Transmission Grating Spectrometer (HETGS). 
The \rxt~data are from a two week follow-up of the source while the  
\cha~observation took place around the end of the follow-up,
about 12 days after the discovery of the source, when the source flux had decreased already by 
a factor of ten.
The analysis of the \cha~data allowed us to extract the most precise X-ray position 
of \igr, RA=$00^{\mathrm{h}}$ $29^{\mathrm{m}}$ $03.08^{\mathrm{s}}$ and Dec=$+59 ^\circ$ 34$^\prime$
19.2$^{\prime\prime}$ (0.6$^{\prime\prime}$ error), compatible with the optical and radio ones. 
We find that the spectra of \igr~can be described by 
a combination of a thermal component and a power-law.
Along the outburst detected by PCA, the power-law photon index shows no particular trend while 
 the thermal component ($\sim$1\,keV, interpreted as a hot spot on the neutron star surface) 
 becomes weaker until non-detection. 
In the simultaneous observation of the weak \cha~/\rxt~spectrum, there is no more indication for
the  $\sim$1\,keV thermal component while  we detect a colder thermal component ($\sim$0.4\,keV) that we interpret 
as the emission from the cold disc.
A hint for a 6.4\,keV iron line is detected, together with  
 an excess around 6.8\,keV and absorption feature around 7.1\,keV. The latter two features  
 have never been detected in the spectra 
  of accretion-driven millisecond pulsars before and, if confirmed, would suggest the presence of 
  an expanding hot corona with high outflow velocities.
  
   }

\authorrunning{A. Paizis et al.}
\titlerunning{\textsl{Chandra} and \textsl{RXTE} Observations of IGR J00291+5934}
\maketitle

%
%________________________________________________________________

%
\section{Introduction}

In Low Mass X-Ray Binaries (LMXRB) hosting a neutron star, the accreting star is thought to be
 very old 
and weakly magnetised, compared to  neutron star High Mass X-Ray Binaries.
It is believed that in LMXRBs, the accretion disc is generally not influenced by the magnetic field and can
 extend very close to the neutron star surface in a slow, ``spot-less'' accretion.
This seems to be the case for the majority of LMXRBs for which no regular 
pulsations have been observed \citep[see e.g.][for a review]{white95,psaltis04}.
In a few cases though, regular X-ray pulsations have been detected with spin periods 
ranging from about 120 sec for \mbox{GX 1+4} to less than 10 msec, e.g. for \sax. 
The discovery of the  latter accretion-powered millisecond pulsar (APMSP)
has been  very important in the LMXRB evolution scenario: the existence of such systems
supports the theory that LMXRBs are indeed the progenitors of millisecond radio 
pulsars with a low magnetic field. 

Currently seven APMSPs are known. A recent review of these systems can 
be found in \cite{wijnands05}\footnote{With the exception of the very recently discovered 
HETE~J1900.1$-$2455 \citep{morgan05}.}. 
Among them, with its 1.67 msec spin period, \igr~is the fastest known APMSP.
Evidence for the existence of rapidly (msec) spinning neutron stars in a wider
sample of LMXRBs (at least in  an additional 11 systems)  has been provided
by the detection of nearly coherent 
oscillations (``burst oscillations'') during type-I X-ray bursts \citep{chakrabarty04}. 
Nevertheless, only for a few LMXRBs can we actually see the spin due to regular X-ray 
pulsations.
The reason for this discrepancy is still debated and it is not clear  why the physics
at the origin of the X-ray pulsations reveals itself only through the presence
 of pulsations and not in the spectral and timing properties of the sources 
that remain similar between APMSPs and non pulsating LMXRBs
\citep{wijnands05}.

In this paper we report on observations of  the fastest discovered  APMSP, 
\igr,  performed with the \cha~{\it X-ray Observatory} \citep{weisskopf02}, 12 days 
after the discovery. To have a better overview of
the spectral behaviour of the source prior to our observation, 
 we have also analysed the available {\it Rossi X-ray Timing Explorer} \citep[\rxt~;][]{jahoda96} 
 observations, part of which were simultaneous to our 18\,ksec \cha~observation.

 The paper is organised as follows: in
section two we give an overview of the current knowledge of  \igr~
from its discovery. In section three we present our observations and data
reduction methods. In section four we present our results that are then 
discussed in the last section. 

\section{\igr }

\igr~was discovered by the {\it INTErnational Gamma-Ray
  Astrophysics Laboratory} \citep[INTEGRAL;][]{winkler03} on December 2nd 2004, during routine
monitoring of the Galactic plane \citep{eckert04}. 
Follow-up observations with {\it RXTE} revealed the presence of coherent
pulsations at $\sim$598.89 Hz \citep{markwardt04} with an energy dependent 
fractional amplitude. 
Further analysis and observations with {\it RXTE}
showed that the neutron star is in a 8844 s (2.45 h) orbit. The upper limit of 0.16 $M_\odot$
on the mass of the companion star in \igr~implies that the companion 
is most probably a hot brown dwarf  \citep{galloway05}.

The outburst X-ray spectrum could be fitted with an 
absorbed power-law with $\Gamma=1.8$ and fixed column density
\nh=2$\times$$10^{21}$\,cm$^\mathrm{-2}$ for \int~\citep{shaw05} and 
 $\Gamma=1.7$ and measured column density
\nh=1.7$\times$$10^{21}$\,cm$^\mathrm{-2}$ for {\it RXTE}  \citep{galloway05}.
More physical models (thermal Comptonisation) led to poorly
constrained electron temperatures with a cut-off between 80--120\,keV 
\citep{shaw05}.
\cite{jonker05} observed \igr~with \cha~one month after the discovery of the source, likely 
at its quiescent flux level, obtaining a 0.5--10\,keV flux of $\sim$$10^{-13}$\ergcms and
a neutron star effective temperature of $\sim$0.3\,keV.

Inspection of the {\it RXTE}/ASM archive showed that the 
source was likely also active in the past on two occasions,
leading to a tentative
recurrence time of about 3 years \citep{remillard04}.

Observations with ground based telescopes  allowed the
counterparts at optical and radio wavelengths to be discovered
\citep{pooley04, fender04, fox04}. The most
accurate optical position of the source is RA=$00^\mathrm{h}$
$29^\mathrm{m}$ $03.06^\mathrm{s}$ and Dec=$+59 ^\circ$ 34$^\prime$
19.0$^{\prime\prime}$ (0.5$^{\prime\prime}$ uncertainty) \citep{fox04}. 

\section{Observations and data reduction}

\begin{figure}
\centering
\includegraphics[width=0.8\linewidth]{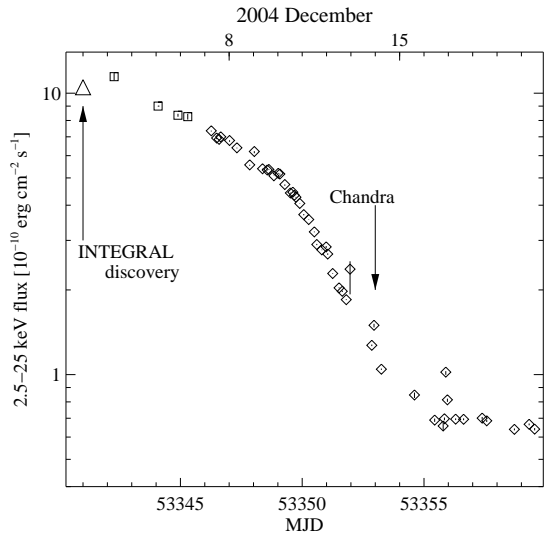}
%\vspace{4cm}
\caption{X-ray intensity of \igr~throughout the 2004 outburst. The open triangle
indicates the discovery with \int~. 
The 2.5--25\,keV PCA flux evolution is shown (open squares).The time of our \cha~
observation is indicated by the arrow in the right. After MJD 53355 the source flux
is strongly influenced by residual flux and variability of (at least) the nearby V709 Cas. 
\label{fig:histo}}
\end{figure}

An X-ray intensity history of the 2004 outburst of \igr~is presented in 
Fig.~\ref{fig:histo},
where the \int~detection followed by the \rxt~follow-up are shown.
The time of our \cha~observation is indicated by the arrow on the right. As can 
be seen, we observed the source at a very faint flux level, 
after the source had decayed significantly 
towards quiescence.

\subsection{The \cha~data}
The 18\,ksec \cha~observation of \igr~was performed  on 2004 December
14th from 02:43:41 until 08:02:38 UT (MJD 53353).
We reduced the data in a standard manner using the CIAO version 3.2 software 
package and CALDB version 3.0.0. The obtained spectra have been analysed with 
the ISIS analysis system, version 1.2.4 \citep{houck02}.

We used the High Energy Transmission Grating Spectrometer, HETGS \citep{canizares00}.
It  has two sets of gratings, the High Energy Grating, HEG, 
and Medium Energy Grating, MEG, covering the energy ranges of 0.8--10\,keV and
0.4--5.0\,keV, respectively. The focal plane imager used is the Advanced CCD 
Imaging Spectrometer (ACIS-S), an array of six CCD detectors normally used as readout
for the photons dispersed by the gratings.
The CCDs were operated in a sub-array mode where only
half the CCD was read out.  This did not affect the dispersed spectra,
but served to reduce the frame integration time from the usual 3.2\,s
to 1.7\,s, and thus reduced the presence of pileup (see below).

Given 
the low signal to noise obtained, we extracted the zeroth-order (undispersed) spectrum and
the first order dispersed spectra ($m=\pm1$ for HEG and MEG) for a total of five spectra.
Higher order spectra were not considered.
The zeroth-order  image of the source is well resolved, but is also
mildly piled up (pile-up fraction of 18\%). 
In all spectral
fits wherein we fit the zeroth-order data, we account for these effects
by using the ISIS implementation of the pileup model of \citet{davis01}.

Within the field of view of ACIS-S we detect another source, 17$^\prime$ away from 
the zeroth-order image of \igr~. The source is V709 Cas, a cataclysmic variable. 
The zeroth-order image of V709 Cas lies on top of the HEG $m=-1$ dispersed spectrum of 
\igr.
The position of V709 Cas corresponds to the  1--1.3\,keV region of the
HEG $m=-1$ spectrum, that has been ignored in the spectral analysis. Outside of
1--1.3\,keV, the removal of events associated with
V709 Cas is extremely efficient and no contamination is occurring.

Inspection of the 5 spectra showed that the separate grating arms agree with each other. 
This allowed us to merge the two HEG ($m=\pm1$) and MEG ($m=\pm1$) spectra into two 
combined spectra 
in order to increase the signal to noise ratio. This led to three final spectra that 
we used in the analysis 
(combined first order HEG, combined first order MEG and zeroth-order).
For the fit, we  binned the data to obtain a minimum of 16 counts per bin for
both HEG and MEG as well as a minimum number of 16 channels per bin for HEG and 8 for MEG.

The 1.67 msec pulsations cannot be detected in our data, due to the
1.7\,s frame integration time in our observations.  We have used both
`unbinned techniques' for searching for variability on all time scales
 $>$ 1.7\,s (e.g., the Bayesian Blocks method of \cite{scargle05}, in
prep.\footnote{See the implementation of this algorithm in ISIS/s-lang
at http://space.mit.edu/CXC/analysis/SITAR/}) as well as `binned'
period folding techniques to search for variability on time scales
comparable to the orbital period.  No statistically significant
variability was found.

\subsection{The \rxt~data}
One day after the discovery of  \igr, a regular monitoring 
of the source was performed with \rxt~via Target of Opportunity (ToO) 
observations starting from December 3rd until December 16th.
The \rxt~observation on December 14th was performed simultaneously to 
our \cha~pointing (from 03:08:48 to 08:53:52 UT).
The \rxt~data were extracted with the HEASOFT software, version 5.3.1
 using our standard procedures \citep{wilms99}. 

The simultaneous \cha/\rxt~fit was performed in the already mentioned ISIS package.
We rebinned further the 
HETGS spectra to obtain a minimum number of channels per bin equal to 64 for HEG and 32 for MEG.
In this way the number of PCA and HETGS data bins is comparable and during the fit similar weight is given to 
all the available spectra (HEG, MEG, PCA).
We fitted the PCA spectra in the 3--13\,keV spectral range with a
 0.5\% systematics and grouped the data to have a 
minimum signal to noise of 5.
The interpretation of the PCA  December 14 spectrum of  \igr~is delicate 
 due to the presence of (at least) the nearby cataclysmic variable V709 Cas 
 (17$^\prime$ away, see Section 3.1) that is clearly detected in the \cha~field of view. 
 
To take into account the contamination from nearby sources, we subtracted
the last \rxt~pointing available from all the previous observations. The integrated 
\mbox{2--8\,keV} flux in this last PCA observation is consistent with the Chandra
determined flux of V709 Cas alone (2.4$\times$10$^{-11}$\ergcms), independently confirming the
non-detection of \igr~in this observation. 

\section{Results}

We extracted the X-ray position of  \igr~from the zeroth-order image, obtaining 
RA=$00^\mathrm{h}$ $29^\mathrm{m}$ $03.08^\mathrm{s}$ and Dec=$+59 ^\circ$ 34$^\prime$
19.2$^{\prime\prime}$ equinox J2000 (90\% confidence error of 
0.6$^{\prime\prime}$\footnote{See http://asc.harvard.edu/cal/ASPECT/celmon/index.html}). 
This position, compatible with the optical and radio ones, 
was immediately announced to the community by \cite{nowak04}.

\begin{figure}[h]
\includegraphics[angle=270,width=0.48\textwidth]{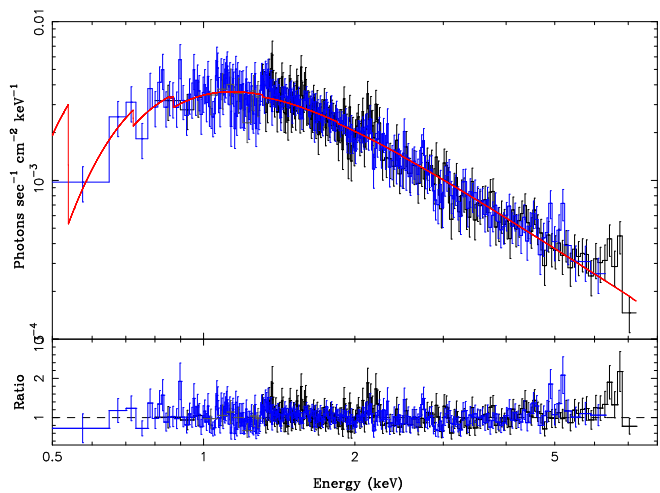}
\hfill
\caption{\cha~HEG (order $\mathrm{m}=\pm$1 combined, black) and 
MEG (order $\mathrm{m}=\pm$1 combined, blue) spectra of \igr~. The model shown is an 
absorbed power-law. 
\label{fig:po}}
\end{figure}

The first order HEG and MEG spectra of \igr~are shown in Fig.~\ref{fig:po}. 
The best fit obtained, also shown in Fig.~\ref{fig:po}, is an absorbed power-law with column density\footnote{The 
improved model for the absorption of 
 X-rays in the interstellar medium "tbvabs" Version 1.0 was used \citep{wilms00}}  
 \nh=(4.3$\pm$0.4)$\times$$10^{21}$\,cm$^\mathrm{-2}$
 and $\Gamma=2.06\pm0.07$ 
with a reduced \chis=0.79 for 431 d.o.f.
The absorbed flux is 2.5$\times$$10^{-11}$\ergcms~in 0.5--8\,keV 
and  1.9$\times$$10^{-11}$\ergcms~in 2--8\,keV ($\sim$1mCrab) with about 10\% uncertainty.

\begin{figure}[h]
\includegraphics[width=0.48\textwidth]{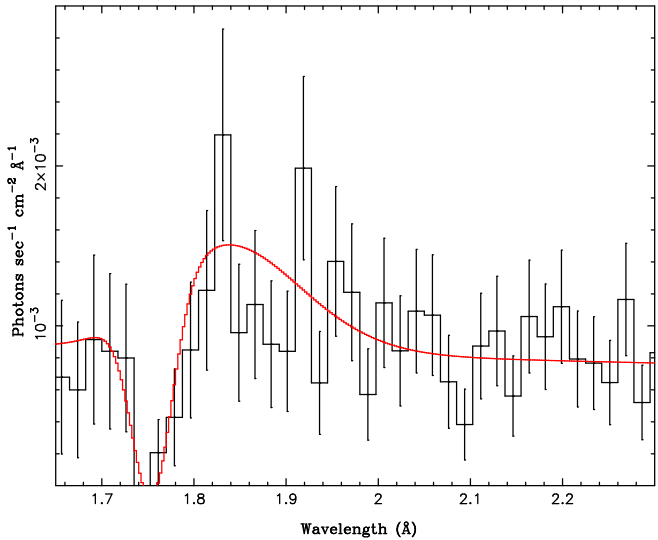}
\hfill
\caption{HEG data  uniformly rebinned by a
factor of 7, where we have fitted this region with a power law plus broad
line plus absorption feature.  The latter is the most significant
feature in our HEG spectra (approximately 99.3\% significant) and is also present in the zeroth-order 
spectrum. 
 \label{fig:hint}}
\end{figure}

Residuals in the continuum are found by taking the best-fit
continuum model, and then using that as a `Bayesian prior' for the
expected count rate in each bin of the \emph{unbinned} spectrum.  One
can then apply the Bayesian Blocks algorithm of Scargle (2005, in
prep.) to search for the most significant data residuals\footnote{This
procedure is described in detail at:
http://space.mit.edu/CXC/analysis/SITAR/bb\_experiment.html}.  Using
this procedure, we find the most significant residuals to be excess
emission between approximately 6.4 and 6.8\,keV (1.94--1.82\,\AA), and an
emission deficit between 7.0 and 7.2\,keV (1.73--1.78\,\AA), as shown in Fig.~\ref{fig:hint}.  
Fitting just the HEG spectrum between 1.5 and 3\,\AA, where we have 
grouped the data uniformly by 7 bins, and using the 
statistic of \cite{cash79} , if we set the normalisation of 
the absorption line to zero and refit, the statistic increases by 10.8. 
  This is approximately the 99.3\% significance threshold.  (Setting the 
normalisation of the emission line to zero and refitting only changes 
the Cash statistic value by 3.)  Thus, barring systematic uncertainties 
of the detector, this absorption feature is approximately 3$\sigma$ 
significant.
Hints for the
presence of these featues are seen in the zeroth-order spectrum as well, as we show below.

The simultaneous HETGS-PCA spectrum gave an overall harder spectrum with 
$\Gamma=1.8$. 
The harder continuum set by PCA  gave a soft excess in the \cha~spectrum, which 
was then adjusted in the fit resulting in a smaller column density (\nh=3.1 $\times$ $10^{21}$\,cm$^\mathrm{-2}$).
Nevertheless the fit was not very good, reduced \chis~of 1.35 (163 d.o.f.), and indeed 
the low end of the MEG spectrum is not well described; the new \nh~is lower than 
the value obtained   fitting the
 \cha~data alone and
predicts more photons in the soft end of the spectrum than what actually detected by the MEG.
Furthermore, the higher end of the PCA spectrum shows a slight excess above 8\,keV, meaning that the 
1.8 slope is not enough for the PCA data.

\begin{figure}[h]
\includegraphics[angle=270,width=0.48\textwidth]{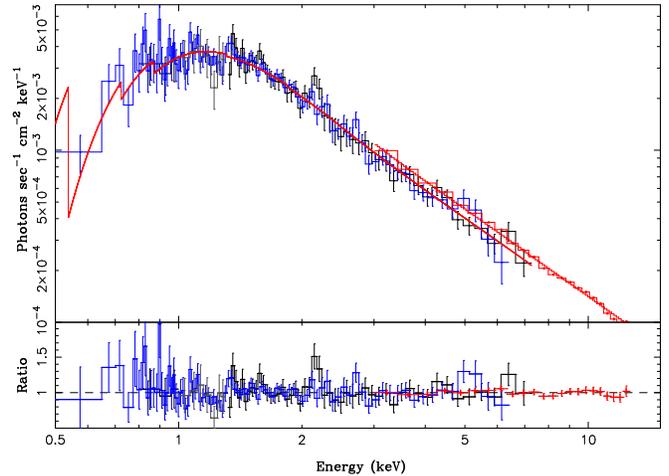}
\hfill
\caption{HEG (black), MEG (blue) and PCA (red) spectra of \igr~. The fit shown is a composition
of an absorbed power-law and disc blackbody.
   \label{fig:all}}
\end{figure}

To adjust for these issues, we  added a soft thermal part to the spectrum \citep[DISKBB model,][]{mitsuda84}
obtaining a very good description of the  data (\chis=0.9, 161 d.o.f.):
the PCA data are better described by the new power-law ($\Gamma$=1.7)
and the softer part of the spectrum is well fitted by \nh=4.3 $\times$ $10^{21}$\,cm$^\mathrm{-2}$
 and  a thermal component of $kT=0.42$\,keV (13\% contribution 
to the absorbed 0.5--8\,keV flux) 
that compensates for the soft excess induced by the hard slope.
Fig.~\ref{fig:all} shows the  HEG, MEG and PCA spectra of
 \igr~together with the best fit model.
  
The discrete feature at 2.1\,keV is a known calibration origin and is due to a jump in 
the HETGS effective area. 
The excess at  $\sim$5\,keV has no known instrumental origin and could be real; however,
it is seen neither in the HEG nor in the zeroth-order spectrum and is less 
significant than the already discussed iron region features that are  mirrored also in the zeroth-order 
spectrum.

We find that the normalisation of the PCA spectrum is about 13\%
higher than for \cha~. Between these two missions, normalisation factor 
up to 20\% 
have been found before \citep[e.g.,][and references therein]{juett03}.
We note here that if we use  a non-cleaned PCA spectrum, i.e. not corrected for the contribution
of the nearby sources, we obtain a normalisation factor of
 50\% with respect to ~\cha. This high value is most likely due to the contamination from  nearby sources
 and background in the PCA spectrum.

\begin{figure}[h]
\includegraphics[angle=270,width=0.48\textwidth]{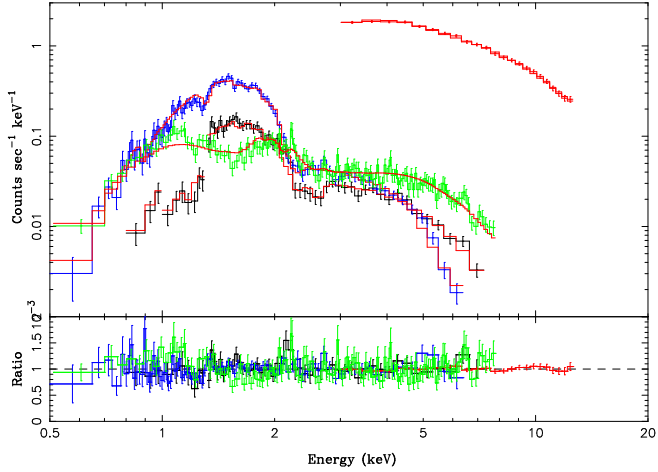}

\hfill
\caption{HEG (black), MEG (blue), PCA (red) and HETGS zeroth-order (green) count spectra of \igr~. 
Note the  emission/absorption line features 
already discussed (Fig.~\ref{fig:hint}) in both the zeroth-order and HEG spectra.
The break in the HEG spectrum (data points between
1--1.3\,keV) is due to the fact that in this range we have used HEG $m=+1$ alone and not the averaged $m=\pm1$
orders, due to the contamination by  V709 Cas in the HEG $m=-1$ order. 
   \label{fig:counts}}
\end{figure}

As a last step, we included the \cha~zeroth-order spectrum that is
affected by pile-up and thus the spectral shape from the source is distorted. 
Fig.~\ref{fig:counts} shows the  HEG, MEG, PCA and zeroth-order count spectra.
The spectral distortion due to the pile-up in the zeroth-order spectrum is visible (dip in 
the 1--2\,keV region in the green spectrum). The emission/absorption  features 
detected by the HEG in the iron region (already discussed, Fig.~\ref{fig:hint}) are 
mirrored in the zeroth-order as well.
The information that is lost due to pile-up can be partly recovered using  kernel models 
that ``correct'' the distortions \cite{davis01}. Nevertheless,
the result cannot be perfect and the inclusion of the zeroth-order
spectrum can make the overall fit quality slightly worse.
We obtain an absorbed power-law and thermal 
component, compatible with the previous case, with column density \nh=3.9$\times$ $10^{21}$\,cm$^\mathrm{-2}$, 
$\Gamma=1.7$, $kT_\text{in}=0.4$\,keV and a reduced \chis=1.15 for 293 d.o.f.
Again we  find a  PCA normalisation of about 13\% higher than for \cha. 
Table ~\ref{tab:all} summarises the results of our spectral analysis.
 
\begin{table}
  \begin{center}
    \caption{Best fit parameters for the absorbed power-law model of the HETGS spectra of \igr~and 
    for the absorbed power-law plus thermal component of the HETGS, PCA and \cha~zeroth-order spectra. 
\nh~is the column density in units of $10^{21}$\,cm$^\mathrm{-2}$;
    $kT_{in}$ is the inner temperature of the thermal (DISKBB) component; 
 d.o.f. = degrees of freedom (the decrease in d. o. f. in the combined HEG / MEG and PCA spectrum is due to the 
 heavy rebinning we performed in the HETGS data, see text); in all the fits we fixed to 1 HEG, MEG and zeroth-order normalisation factors.
  The indicated errors are at 1$\sigma$. \label{tab:all}}
    %\vspace{1em}
    \renewcommand{\arraystretch}{1.2}
    \begin{tabular}[h]{lll}
      \hline
      \hline
 &  HEG / MEG & HEG / MEG / PCA \\
      \hline
\nh & 4.3$\pm$0.4& $4.3 {+0.7 \atop -0.5}$ \\
$\Gamma$ &  2.06$\pm$0.07 & 1.7$\pm$0.08\\
$kT_{in}$ &    & $0.42 {+0.07 \atop -0.06}$\,keV \\
PCA norm & & 1.13\\
 reduced \chis & 0.79 (431 d.o.f.) & 0.9 (161 d.o.f) \\
\hline
\hline
 & HEG / MEG & \\
 & PCA  / 0th order& \\
\hline
\nh & $3.9{+0.5 \atop -0.4}$ & \\
$\Gamma$ & 1.7$\pm$0.04& \\
$kT_{in}$ & 0.40$\pm$0.06\,keV & \\
PCA norm & 1.13 &\\ 
reduced \chis & 1.15 (293 d.o.f.) & \\
 \hline
    \end{tabular}
    \label{tab:grating}
  \end{center}
\end{table}

To have an overview of the outburst behaviour of \igr~before our \cha~observation, we 
analysed all the available \rxt/PCA observations. The PCA spectra of \igr~can be  fitted 
by a combination of a thermal component ($\sim$1\,keV) and power-law
($\Gamma$$\sim$1.6)\footnote{This slope is adjusted to the softer 1.7 value when fitted with the \cha~
simultaneous spectrum of December 14.}. A hint for an iron line at 6.4\,keV (fixed) is detected.
During the outburst, while the source is decaying, the power-law photon index shows no particular trend 
(unlike the normalisation that decreases significantly) while 
 the thermal component  becomes weaker and can be constrained no 
more by PCA. Already two days before our \cha~observation the \rxt~observations are consistent 
with a single power-law with a marginal detection of the 6.4\,keV line. 
We believe that in our \cha~spectrum we still see a residual of the disappearing 6.4\,keV line
as well as a soft component (0.4\,keV) that cannot be constrained by PCA alone.

\section{Discussion}

We have studied the spectral evolution  of the accretion-powered millisecond pulsar \igr~with 
\cha~and \rxt~during the December 2004 outburst that led to its discovery by \textit{INTEGRAL}. 

\subsection{Long term variability}

At its discovery \igr~had a 5--100\,keV flux of 
about 10$^\mathrm{-9}$\ergcms \citep{shaw05} that decayed 
to about 10$^\mathrm{-10}$\ergcms (extrapolated from our HETGS/PCA best fit) about
12 days later.
Assuming a distance of 5\,kpc \citep{galloway05} we obtain that the X-ray luminosity changed 
from  L$_{X}$$\sim$10$^\mathrm{36}$\ergs to L$_{X}$$\sim$10$^\mathrm{35}$\ergs. 
This corresponds to a change from $\sim$0.01 to  $\sim$0.001\,$L$$_{Edd}$ that places
\igr~at the lower end of the dim Atoll sources ($\sim$0.01--0.3\,$L_{Edd}$), 
similarly to other APMSPs.
Such a peak luminosity is very low if we consider the bright
neutron star X-ray transients (XRTs) that display  outbursts 
with peak luminosities of 
L$_{X}$$\sim$10$^\mathrm{37}$-10$^\mathrm{38}$\ergs. 
 The history lightcurve of  \igr~from the \rxt/ASM indicates that the flux
measured at the time of the discovery is indeed the peak of the outburst but
unfortunately, the distance to \igr~is not well constrained and we cannot
be sure of the absolute value. It may seem reasonable to
expect that the peak luminosity is indeed low: APMSPs are very compact systems, 
 the disc outer radius R is relatively small 
compared to wider orbit LMXRBs and this limits the total disc mass 
($\propto$R$^3$) 
that can build up in quiescence.  When the outburst is triggered,
the mass accretion onto the central object is $\propto$R$^{2}$
\citep[][and references therein]{gierlinski02b}.  Nevertheless, 
things are not so straightforward and there are cases where the above scenario does not hold,
e.g. in the case of  the ultra compact (11.4 minute period) LMXRB 
4U~1820$-$30 that can still reach X-ray luminosities higher 
than a few times 10$^\mathrm{37}$\ergs~\citep{ballantyne04}.

APMSPs differ from the bright XRTs also in the quiescent luminosity: 
standard XRTs range between \mbox{L$_{X}$$\sim$10$^\mathrm{32}$-10$^\mathrm{33}$\ergs}
\citep{campana02} while APMSPs seem to be dimmer than $\sim$10$^\mathrm{32}$\ergs
\citep{wijnands05}. 
Our \cha~measurement of \igr~gives a 0.5--8\,keV flux of $\sim$$10^{-11}$\ergcms
that, compared to the 0.5--10\,keV flux of $\sim$$10^{-13}$\ergcms by \cite{jonker05}, 
clearly shows that we detected \igr~during its outburst decay, prior to its
quiescent state.

\subsection{The spectral continuum}
The study of the other APMSPs has shown that during the outburst
the spectra can be well fitted by a two component model, a soft thermal component and a 
harder one. In general the soft component is interpreted as blackbody emission 
from a heated hot spot on the neutron star, responsible for the X-ray pulsations, 
and ranges within about 0.6--1\,keV \citep[see e.g.][]{gierlinski02b,gierlinski05,juett03}.
The harder component is interpreted as thermal Comptonization by a plasma
heated by the accretion shock as the material, collimated by the magnetic field, 
impacts onto the neutron star surface. The source of seed photons for 
Comptonisation seems to be either the hot spot itself \citep{gierlinski05}
 or the colder accretion disc \citep{titarchuk02}.

In our analysis, we find that the spectra of \igr~can be  fitted 
by a combination of a thermal component and a power-law.
Along the outburst, while the source is decaying, the power-law photon index shows no particular trend while 
 the thermal component becomes weaker and can be constrained no 
more by PCA. At this point, we obtained our \cha~observation and the 
simultaneous weak \cha/\rxt~spectrum can be well described by 
the combination of a colder thermal component ($kT_{in}=0.4$\,keV instead of $\sim$1\,keV) 
and a power-law. 
The presence of the power-law  means that if there is a cut-off in the spectrum, then this occurs at higher energies 
than the HETGS/PCA range and we are not able to constrain it. A Comptonisation 
model in our data gives a poorly constrained Comptonising plasma temperature of about 50\,keV. 
The presence of the thermal component means that not all the available soft photons
go through the Comptonising medium, instead there are parts that are seen
 directly, similarly to what was found in other APMSPs in outburst, 
\cite{miller03} and \cite{gierlinski05} on \mbox{XTE J1751$-$305}, 
\cite{gierlinski02b} on \mbox{SAX J1808.4$-$3658} and 
\cite{juett03} on \mbox{XTE J0929$-$314}.

We interpret the evolution we see in PCA data and the final simultaneous PCA/\cha~spectral properties 
in the following way: at the on-set of the outburst
the accreting matter is channelled onto the neutron star surface and a hot spot is created. We have indications for
the hot spot  in the early PCA data with a thermal component around 1\,keV that is likely to be responsible 
for the pulsations detected by \cite{galloway05}. 
The disc is colder ($<$1\,keV) and we do not detect it directly in the PCA spectra alone, but
we see its effect in a hint for the  6.4\,keV line in the PCA data, most likely originating from irradiation of the cold 
accretion disc by the X-ray source.\footnote{A direct simultaneous measure of the hot spot and disc has been reported 
in the case of the APMSP XTE J1751$-$305 
by \cite{gierlinski05} where two soft components were detected in the broad-band 
\textit{XMM-Newton}/\rxt~spectrum: $\sim$0.6\,keV, associated to the disc and $\sim$1\,keV, the hot spot.}  
Along the outburst, the hot spot 
becomes progressively weaker and already 10 days after the on-set of the outburst we are not able to 
constrain it with PCA anymore. This is consistent with the fact that \cite{galloway05} detect no pulsations 
in the last phases of the outburst. The disc is likely becoming colder as it is illuminated by the fading 
X-ray source and is possibly receding after the outburst, similarly to other LMXRBs in which at 
a low accretion rate, the disc is more distant from the compact object and the overall spectrum is hard.
At the time of our \cha~observation the hot spot (along with its pulsations) is most likely off and we
detect a cold accretion disc around 0.4\,keV. 

\subsection{Discrete features in the spectrum}
Despite the regular search for discrete features in APMSPs in the past, no line was found besides the 
 6.4\,keV one in \sax~\citep{gierlinski02b}.
In our study,  a hint for an iron K$\alpha$ line (fixed at 6.4\,keV) is detected in the PCA observations, while 
 a marginally significant excess in the 6.4--6.8\,keV region and a 3$\sigma$ significance deficit 
 around 7.1\,keV are detected in both the HEG and zeroth-order \cha~spectra (Figs.~\ref{fig:hint} and~\ref{fig:counts}).  
Emission lines at 6.4\,keV have been observed from many X-ray pulsars and are believed to be produced 
by fluorescence of weakly ionised iron  
(less than Fe XV). A likely candidate for this cold matter is the accretion disc that 
can be also responsible  for the iron K-edge $\sim$7.1\,keV feature we detect in the spectrum.
The emission line that we detect around 6.8\,keV (if real) would require the presence of a highly ionised,
i.e. hot, corona in \igr, consistent with the high plasma temperature ($>$10\,keV)
inferred from the spectral study performed here, as well as in \cite{shaw05}. This scenario 
would require a different medium for each feature of Fig.~\ref{fig:hint} (cold disc for the 7.1\,keV feature
and hot corona for the 6.8\,keV one). 

Another possible, although speculative, interpretation of the discrete features in Fig.~\ref{fig:hint}
could be the following: 
 the emission feature we detect around 6.8\,keV could be the
redshifted part of the 6.97\,keV line (expected from Fe XXVI) while  the absorption feature 
around 7.1\,keV could be the blueshifted 6.97\,keV  line.  In this scenario we 
would basically obtain that the  6.4\,keV line is produced by the cold disc while the 6.8 and 7.1\,keV lines
are a P-Cygni profile from an expanding hot corona with outflow velocities of about 6000\,km/sec.
Such (and higher) outflow velocities are possible and have already been observed in 
cataclysmic variables \citep[e.g.,][]{woods92}
and in the X-ray binary SS433 \citep{migliari05}. The P-Cygni profile scenario is consistent with all the available data we
currently have on \igr. However, the relatively low statistical significance of the discrete features does not allow us to firmly 
establish this interpretation and underlines the importance of obtaining quick follow-ups of these transient events.

\begin{acknowledgements}
We would like to thank the \cha~team for the help during the trigger of the ToO and Jean Swank and Evan Smith
for performing the \rxt~observation simultaneous to our \cha~one. AP thanks J. Poutanen for sharing his
overview knowledge on the physics of APMSPs and S. Shaw for careful reading of the manuscript.
 AP and PU acknowledge the Italian Space Agency financial 
and programmatic support via contracts ASI-I/R/046/04. MN was supported by NASA grant SV3-73016. This 
work was partly supported by NASA Grant GO4-5049X.

\end{acknowledgements}
\bibliographystyle{aa}
\bibliography{biblio}

\end{document}